\documentclass[3p,times,twocolumn]{elsarticle}

\usepackage{ecrc}

\volume{00}

\firstpage{1}

\journalname{Nuclear Physics B Proceedings Supplement}

\runauth{}

\jid{nuphbp}

\jnltitlelogo{Nuclear Physics B Proceedings Supplement}

\usepackage{amssymb}

\usepackage{lineno}

\usepackage[figuresright]{rotating}

\begin{document}

\begin{frontmatter}



\dochead{}

\title{Jiangmen Underground Neutrino Observatory}


\author{Miao He}
\ead{hem@ihep.ac.cn}

\address{Institute of High Energy Physics, Beijing}
\address{On behalf of the JUNO collaboration}

\begin{abstract}
The Jiangmen Underground Neutrino Observatory (JUNO) is a multipurpose neutrino-oscillation experiment designed to determine the neutrino mass hierarchy and to precisely measure oscillation parameters by detecting reactor antineutrinos, observe supernova neutrinos, study the atmospheric, solar neutrinos and geo-neutrinos, and perform exotic searches, with a 20 kiloton liquid scintillator detector of unprecedented 3\% energy resolution (at 1 MeV) at 700-meter deep underground and to have other rich scientific possibilities. Currently MC study shows a sensitivity of the mass hierarchy to be $\overline{\Delta\chi^2}\sim 11$ and $\overline{\Delta\chi^2}\sim 16$ in a relative and an absolute measurement, respectively. JUNO has been approved by Chinese Academy of Sciences in 2013, and an international collaboration was established in 2014. The civil construction is in preparation and the R\&D of the detectors are ongoing. A new offline software framework was developed for the detector simulation, the event reconstruction and the physics analysis. JUNO is planning to start taking data around 2020.
\end{abstract}


\end{frontmatter}


\section{Neutrino mass hierarchy}
\label{}
The neutrino mixing angle $\theta_{13}$ was determined to be non-zero recently by Daya Bay~\cite{dyb.prl} and other reactor and accelerator neutrino experiments. The unexpected large value of $\theta_{13}$ makes it easier to determine the neutrino mass hierarchy (i.e., sign of the mass-squared difference $\Delta m^{2}_{31}=m^{2}_{3}-m^{2}_{1}$, where $\Delta m^{2}_{ij}$ represents the mass difference of two neutrino mass eigenstates $m^{2}_{i}$ and $m^{2}_{j}$) and the leptonic CP-violating phase $\delta$. The mass hierarchy information can be extracted from the matter-induced oscillation probability of long-baseline accelerator neutrinos and atmospheric neutrinos. It can also be determined by precisely measuring the energy spectrum of reactor antineutrinos at a medium baseline of $\sim$50 km, and by looking for the interference between two oscillation frequency components driven by $\Delta m^{2}_{31}$ and $\Delta m^{2}_{32}$, respectively~\cite{mh1}\cite{mh2}. A Fourier transform of the $L/E$ spectrum, where $L$ is the baseline and $E$ is neutrino energy, can enhance the information of oscillation frequencies and thus improve the sensitivity of the mass hierarchy~\cite{zhanl.fourier1}\cite{fourier}. A standard $\chi^2$ fitting was also played taking into account systematics and the impact of additional constrain from $\Delta m^{2}_{\mu\mu}$ (which is an approximation of $\Delta m^{2}_{31}$ or $\Delta m^{2}_{32}$)\cite{liyf.chi2}. Such a measurement requires a very strict experiment condition including the energy resolution, energy nonlinearity, the baseline and event statistics~\cite{zhanl.fourier2,xqian}.



\section{The JUNO experiment}
\begin{figure}[h]
\begin{center}
\includegraphics[width=\columnwidth]{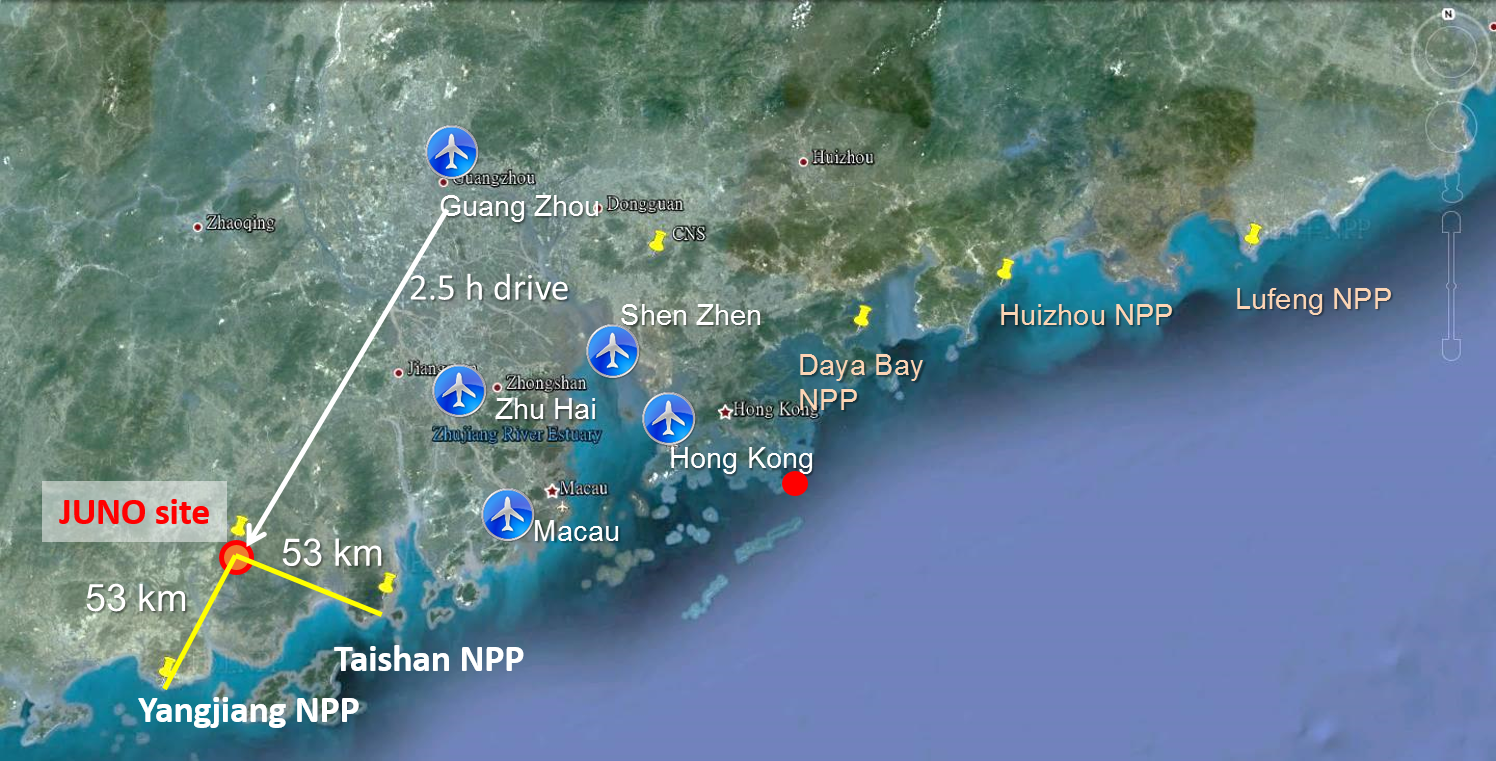}
\caption{\label{fig:juno_site} The experiment site of JUNO. It is located in Kaiping, Jiangmen, Guangdong Province, China, 53~km far from Yangjiang and Taishan nuclear power plants.}
\end{center}
\end{figure}

The Jiangmen Underground Neutrino Observatory (JUNO) is a multipurpose neutrino-oscillation experiment whose major goals are determining the neutrino mass hierarchy and precisely measuring three of the neutrino oscillation parameters. As shown in Figure~\ref{fig:juno_site}, it is located at 700-meter deep underground and is 53~km far from two under-constructed nuclear power plants (NPPs), Yangjiang and Taishan, with a planned total thermal power of 36~GW. There is no other nuclear power plant within 200~km. A 20~kton liquid scintillator (LS) detector with $\sim$15,000 20-inch high detection efficiency photomultiplier tubes (PMTs) is planned to reach $\sim$80\% optical coverage thus to achieve unprecedented 3\% energy resolution (at 1 MeV). A water pool protects the central detector from natural radioactivity in the surrounding rocks. It also serves as a water Cherenkov detector after being equipped with $\sim$1,500 20-inch PMTs, to tag cosmic muons. There is another muon tracking detector on top of the water pool, used to improve the muon detection efficiency and to get better muon tracking.

The reactor electron antineutrino interacts with hydrogen via the inverse $\beta$-decay (IBD) reaction in the LS, and releases a positron and a neutron. The positron deposits its energy quickly, providing a prompt signal. The neutron is captured by a proton after an average time of $\sim$200~$\mu$s, then releases a 2.2~MeV gamma, providing a delayed signal. The coincidence of prompt and delayed signals provides a distinctive antineutrino signature. The estimated rate of IBD candidates is $\sim$65/day assuming full reactor power and 80\% detection efficiency. The dominate background is the correlated $\beta$-n decay from cosmogenic $^9$Li/$^8$He isotopes, which is estimated to be 1.8/day with an anti-coincidence of 4~m in space and 1.5~s in time with respect to a muon track. Other backgrounds such as the accidental coincidences, the cosmogenic energetic neutrons and the $^{13}C(\alpha,n)^{16}O$ reactions are either very small or can be precisely measured in data.

The sensitivity of the mass hierarchy has been studied with simulation by assuming a 20~kton LS target, $3\%/$$\sqrt{E(MeV)}$ energy resolution, 36~GW of reactor power and six years of running time. The real distribution of ten reactor cores in Yangjiang NPP and Taishan NPP is taken into account, and the average baseline is $\sim$53~km with the maximum difference of the baselines of 720~m. The remote reactors in the Daya Bay NPP and the possible Huizhou NPP are also included. A standard Gaussian $\chi^2$ function is defined with a set of pull parameters accounting for the systematic uncertainties including the absolute and relative reactor uncertainty(2\% and 0.8\%), the flux spectrum uncertainty (1\%) and the detector-related uncertainty (1\%). The energy nonlinearity can be self-calibrated by fitting to data since there are multiple oscillation cycles in the $L/E$ spectrum each of which carries the same information of $\Delta m^{2}$, thus the impact on the sensitivity was found to be negligible~\cite{liyf.chi2}. A comparison of the minimal $\chi^2$ when fitting to both normal and inverted mass hierarchies gives a sensitivity of $\overline{\Delta\chi^2}\sim 11$ level. The constrain of $\Delta m^{2}_{\mu\mu}$ with a sub-percent precision by the future accelerator experiments can improve the sensitivity to $\overline{\Delta\chi^2}\sim 16$ level.

%

\section{Recent progress in JUNO}

The JUNO project has been approved by Chinese Academy of Sciences under the Strategic Priority Research Program on February 1, 2013. The geological survey was completed in 2013. The contract of the engineering design, purchase and construction was signed in April 2014 and the land was delivered afterwards. The civil engineering design is nearly finished and there is going to be a groundbreaking ceremony at the experiment site in January 2015. The expected civil construction period will be three years.

\begin{figure}[htb]
\centering
\includegraphics[width=0.9\columnwidth]{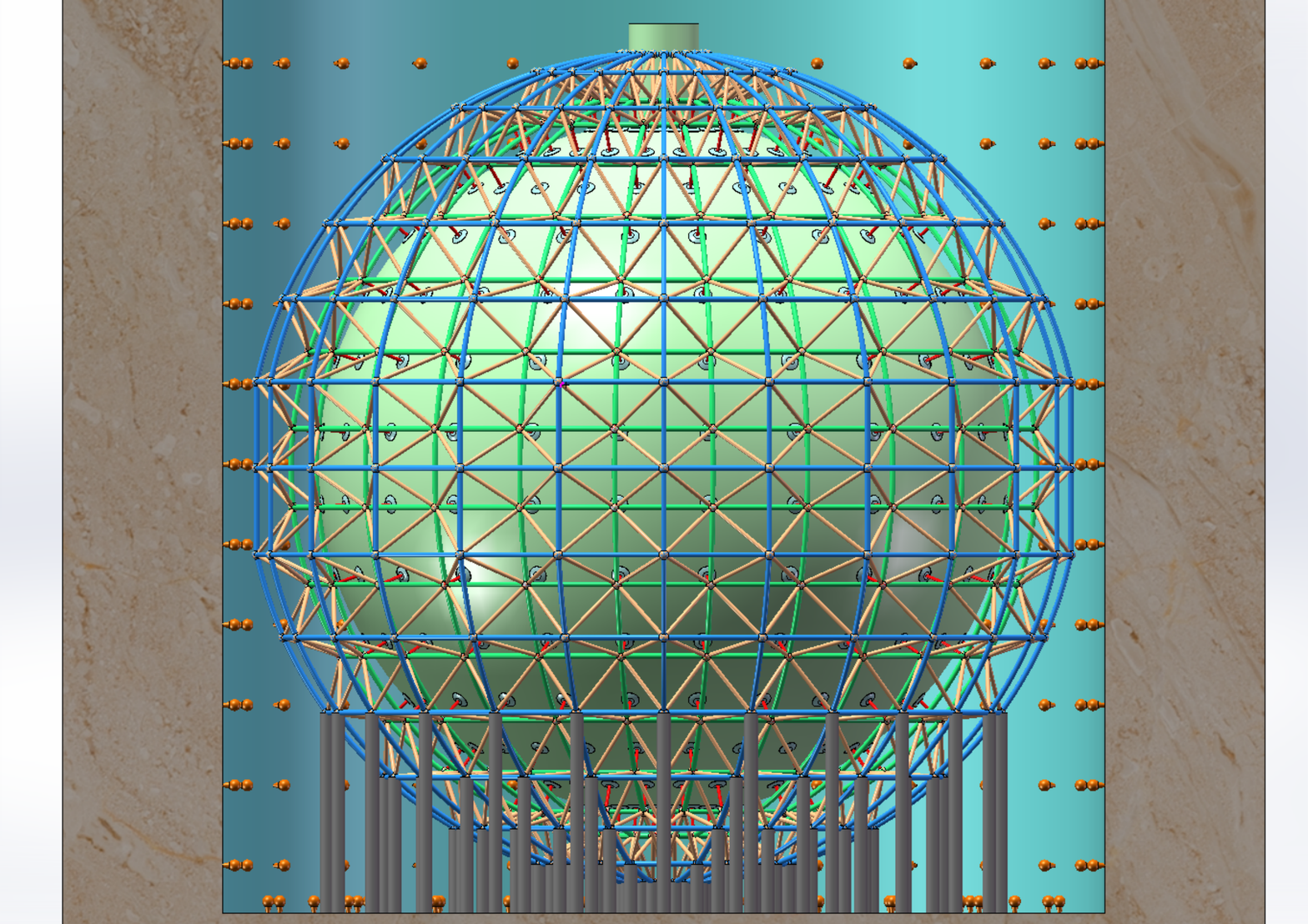}
\caption{\label{fig:detector_default} The default option of the central detector: an acrylic sphere and a stainless steel truss. It is contained in a water pool equipped with PMTs as a water Cherenkov detector.}
\end{figure}

The central detector consists of an inner transparent sphere about 35.5~m in diameter filled with LS and an outer support structure 38~m in diameter. The construction of such a large detector, the test and installation of about 15,000 PMTs, filling and the long-time cycle of different liquid in the detector are very challenging. Multiple detector options were designed and two of them survive up to now. The default option consists of an acrylic sphere as the inner vessel and a stainless steel truss as the support for both the acrylic and PMTs, as shown in Figure~\ref{fig:detector_default}. Finite element analysis has been done including the stress, the deflection and general stability in different loading conditions. The effect of the seismic load, the temperature or the liquid level has also been analyzed. Small components of the detector have been made to test the supporting structure and to study the performance of the acrylic material. In the backup option, a balloon is designed as the container of LS, supported by an acrylic ball. Both of them have the same diameter and are inside a stainless steel tank, which is filled with linear alkylbenzene (LAB) or mineral oil as buffer.

3\% energy resolution at 1 MeV corresponds to 1,200 photon electrons per MeV, which is a much better performance than the state of the art detector such as BOREXINO\cite{borexino} or KamLAND\cite{kamland}. The technical challenges are to make a new type of PMT with high efficiency and to obtain highly-transparent LS. The R\&D effort to develop PMTs for JUNO started in early 2009 in China. A new concept was proposed to make a spherical PMT with the top hemisphere used as transmission photocathode and the bottom hemisphere as reflective photocathode. The conventional dynode is replaced by a back-to-back pair of micro channel plates (MCPs) which has near 4$\pi$ acceptance, thus largely increases the photon collection efficiency. Together with the improvement of the material of photocathode, the total photon detection efficiency is expected to be larger than 30\% in a rather broad spectrum. 5-inch and 8-inch MCP-PMT prototypes were made and tested firstly, and a few 20-inch prototypes were made in summer 2014. The LS for JUNO is composed of LAB, PPO and bis-MSB, without gadolinium doping comparing to Daya Bay, to get lower radioactivity and higher transparency. Current R\&D effort focuses on the purification of the raw material such as vacuum distillation and column filtration. The attenuation length of a small LAB sample made by Nanjing LAB factory reaches to 20~m and 25~m before and after purification, respectively. The characterization of the LAB and LS is ongoing by the measurements of attenuation length, light yield, impurity, Rayleigh scattering and the energy response.

Current design of the veto system includes a water Cherenkov detector, the circulation purification system, a top tracker using plastic scintillator from OPERA\cite{opera} as the baseline option, the geomagnetic field shielding system and the mechanical system. Other systems including detector calibration, readout electronics, trigger, data acquisition, detector monitoring and control are also being designed.

A new offline \textbf{S}oftware framework has been developed for \textbf{N}oncoll\textbf{i}der \textbf{P}hysics \textbf{E}xpe\textbf{R}iments named as SNiPER. It is designed to improve the flexibility and the efficiency of the simulation and analysis by the implementation of a flexible event buffer, the lazy-loading of cascade data objects, and the minimal requirement to external libraries. The full chain of the simulation and the reconstruction has been implemented in SNiPER. MC studies shows a 3\% energy resolution at 1~MeV for both detector options using the optical parameters from Daya Bay except for the change of PMT detection efficiency to 35\% and the LS attenuation length to 20~m.

\section{Summary and perspective}
JUNO was proposed a few years ago (known as Daya Bay II), now boosted by the large $\theta_{13}$. The project has been approved and the collaboration was established. The civil construction is going to start in January 2015. Detector R\&D is ongoing and the offline software is under development. JUNO plans to take data since 2020 and is expected to be running for 20 years with rich scientific goals.

This work is supported by the Strategic Priority Research Program of the Chinese Academy of Sciences, Grant No. XDA10010100, XDA10010900.







\end{document}